\documentclass[%
 reprint,
 amsmath,amssymb,
 aps,
]{revtex4-1}

\usepackage{graphicx}% Include figure files
\usepackage{dcolumn}% Align table columns on decimal point
\usepackage{bm}% bold math
\usepackage[table,xcdraw]{xcolor}
\begin{document}

%\preprint{APS/123-QED}

\title{Engineering Nanoscale Biological Molecular Motors}% Force line breaks with \\

\author{Chapin Korosec}
 %\altaffiliation{ckorosec@sfu.ca}%Lines break automatically or can be forced with \\
\author{Nancy R. Forde}%
 \email{nforde@sfu.ca}
\affiliation{%
Department of Physics, Simon Fraser University, \\
Burnaby, BC, V5A1S6 Canada  
}%


%\affiliation{
%\affiliation{
%\affiliation{%

\date{\today}% It is always \today, today,

\begin{abstract}
\begin{description}

\item[Summary]
Understanding the operation of biological molecular motors, nanoscale machines that transduce electrochemical energy into mechanical work, is enhanced by bottom-up strategies to synthesize novel motors. 
\end{description}
\end{abstract}

\pacs{Valid PACS appear here}% PACS, the Physics and Astronomy
\maketitle

%\tableofcontents

\section{\label{sec:level1}INTRODUCTION}

In his book `What is life?', Erwin Schr{\"o}dinger equated death with the living cell decaying into thermal equilibrium with its surroundings. That is, when a cell is unable to continue using the free energy of its environment to keep order, it decays into a state of disorder. Evolution has led to incredibly beautiful, complex, and astonishing constructs called molecular motors to maintain order and sustain life.

Molecular motors are tiny, nanometer-sized, protein-based machines often described as the `workhorses' of the cell. They are involved in a wide range of key processes, including intracellular transport, positioning of cell organelles, cell division, muscle contraction, and synthesis of ATP, the chemical `fuel' of the cell (see table 1 for biochemical terminology). 

In an effort to understand the complex operation of biological molecular motors, researchers have sought to create synthetic ones. The importance of this approach, a young and growing area of research, was recognized by this year's Nobel Prize in Chemistry. Following the view of Richard Feynman, ``What I cannot create, I do not understand'' [1], we test our understanding of mechanisms of operation through designing artificial analogues of the processes we observe in Nature.

In this article we highlight some properties of biological molecular motors, then focus on the relatively new field of biologically inspired \textit{artificial} molecular motors.

\section{\label{sec:level1}KINESIN: A PROTOTYPICAL WALKING MOTOR}

A well known family of biological motors is kinesin. The motor protein kinesin I functions primarily to transport large cargo that will not randomly diffuse in a reasonable time scale to its needed place in the cytoplasm. It does so by walking in a hand-over-hand fashion directionally along a microtubule `track' within the cell. Kinesin I consists of two flexibly linked globular head groups (the `feet') each with 2 binding sites: one site for ATP and the other for the microtubule surface (\textbf{Figure 1: top}). When kinesin's leading head is bound to a microtubule, ATP can bind into the active site. The neck linker then stiffens such that the lagging head is propelled forward to become the new leading head [2]. Random thermal fluctuations ultimately bring the leading head to dock to the next microtubule binding site. The ATP in the lagging head is hydrolyzed to ADP + phosphate. Rapid diffusion of the phosphate ion into solution results in a sub-nanometre `gap' within the ATP active site [3]. It is believed that this triggers a conformational reorganization near the active site that is mechanically transmitted to the microtubule binding site. This lagging motor head then unbinds from the track, and the cycle repeats itself (\textbf{Figure 1: bottom}). Interestingly, there is evidence that Kinesin I remains in an idle position until a cargo binds [4]; the binding of cargo activates the motor to begin moving!

Kinesin by the numbers: Kinesin I takes 8 nm steps at an average velocity of 750 nm/s, and is able to exert a force of up to 5.4 pN [5]. Given that kinesin's mass is 184 kDa (where 1 kDa = 1.66$\mathrm{x}10^{-9}$ picograms), its ratio of maximum force to body weight is over $\mathrm{10^{9}}$. If a person were capable of such a force-to-body weight ratio they would be able to lift more than 40,000 Boeing 747 aircraft. The incredible difference between the relative capabilities of kinesin versus people arises because molecular systems operate in a stochastic environment where they harness energy in the form of thermal fluctuations (for a detailed discussion on this topic see the article by David Sivak and Aidan Brown in this issue).  

\begin{figure}
	\includegraphics[width=0.5\textwidth]{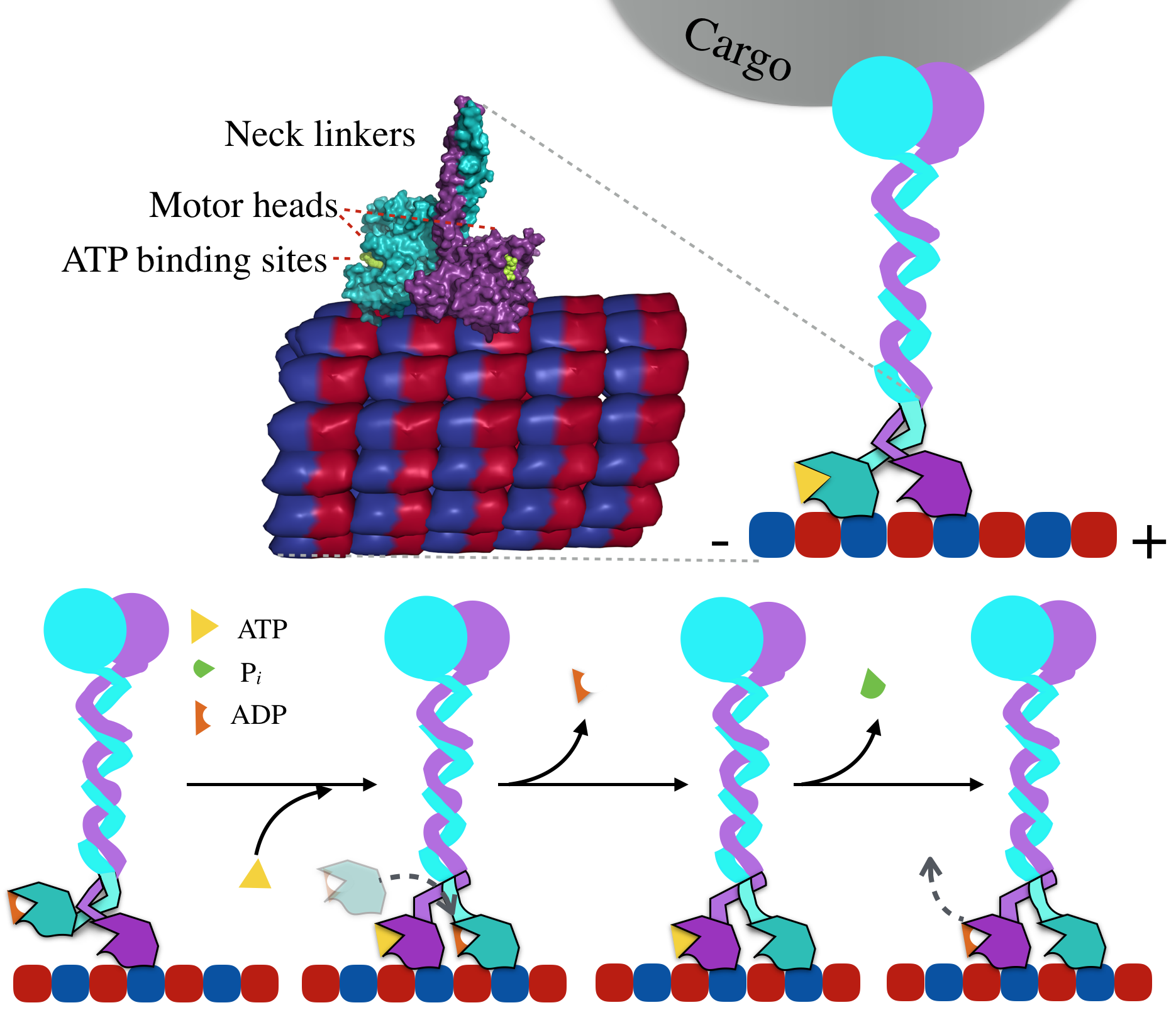}
	\caption{\textbf{Kinesin crystal structure \& walking schematic}. Top: crystal structure of kinesin I motor heads (Protein Data Bank (PDB) 1N6M) and neck linkers docked onto a microtubule (PDB 3JAO). Kinesin motor heads bind tubulin (the single protein building units of microtubules) and walk towards the plus end of the microtubule (towards the cell periphery). Bottom: A motor head in the empty state binds the microtubule, with the lagging head behind. The neck linker stiffens in response to ATP binding, which drives the lagging head forward towards the next binding location. ATP in the now lagging head is hydrolysed to ADP and phosphate (P$\mathrm{_{i}}$), which results in the detachment of this head from the track. The cycle continues with each ATP hydrolysed corresponding to one step forward.} 
\end{figure}

\section{\label{sec:level1}SYNTHETIC WALKING MOTORS}

Inspired in part by kinesin, DNA-based nanomachines have shown great promise as artificial molecular motors. Here, there are two sources of energy that can be used to bias motion: hydrogen bonding associated with DNA basepairing, and hydrolysis (cleavage) of DNA's phosphodiester backbone [6]. The free energy change of creating 10 base pairs of DNA is approximately -58 kJ $\mathrm{mol^{-1}}$ (-23 k$_{B}$T at 300K), similar to that of ATP hydrolysis at standard conditions, -32 kJ $\mathrm{mol^{-1}}$ (-13 k$_{B}$T at 300K)[6]. For reference, the thermal energy scale of k$_{B}$T is 2.5 kJ/mol (or $\sim$0.03 eV) at 300 K. As an example of a DNA nanomachine, Tosan Omabegho and co-workers [7] successfully constructed a bipedal DNA walker that relies on diffusion to move directionally on an asymmetric track. Its hand-over-hand translocation (between DNA footholds supported on a double-stranded DNA track) is analogous to the motility mechanism of kinesin. Its anticipated speed is, however, much slower than that of kinesin due to the timescales of DNA hybridization and unbinding. This DNA nanomachine is one of the few autonomous artificial molecular motors thus far realized.

Another approach to artificial design, much more in its infancy, utilizes non-motor-protein modules as building blocks. The `tumbleweed'[8] is an example of a tripedal motor designed to move on a periodically patterned, linear dsDNA track. Its feet are three unique DNA-binding protein domains each chemically linked to a central hub. By flowing through the appropriate ligand for each foot [9], the construct is expected to take successive steps along a DNA track in a tumbling fashion, as shown in \textbf{Figure 2}. Similar to most DNA motors, the tumbleweed cannot operate autonomously, but relies on rectified diffusion, controlled by cyclically flowing in the correct activators for each foot. For this design, the free energy source for biasing movement is a temporal change in chemical potential: the high concentration of new ligand drives a specific foot to bind, and when that ligand is removed from solution the low concentration leads to the release of the foot. 

\begin{figure}
	\centering
	\includegraphics[width=0.5\textwidth]{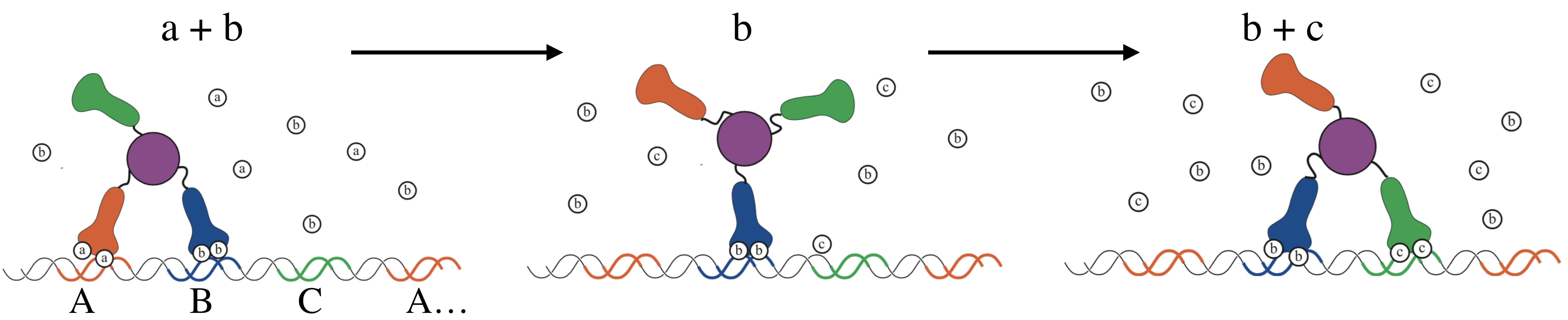}
	\caption{\textbf{Protein-based design of the tumbleweed.} Figure adapted from ref. [8]. The tumbleweed motor is comprised of three chemically linked DNA-binding proteins. The DNA track is patterned with binding sequences specific for each protein  on the tumbleweed motor. The presence of the corresponding ligand for each foot facilitates its binding. From left to right: The tumbleweed begins with the \textit{A} and \textit{B} feet bound in the presence of ligands \textit{a} and \textit{b}. When the solution is replaced by one containing only ligand \textit{b}, the resulting lower concentration of ligand \textit{a} leads to dissociation of ligand \textit{a} from foot \textit{A} resulting in the release of the foot from the track. Ligands \textit{b} and \textit{c} are then flowed in, which promotes the additional binding of foot \textit{C}.  The process continues cyclically to step the motor along the one-dimensional DNA track.} 
\end{figure}

\section{\label{sec:level1}BURNT-BRIDGES RATCHETS: BIOLOGICAL AND SYNTHETIC}

Another class of motile proteins in biology achieves directional and processive motion through a `burnt-bridges' mechanism. A burnt-bridges walker destroys track binding sites as it progresses, thereby biasing its own motion by preventing backwards stepping. 

Matrix metalloproteases (MMPs) are an example of such a system [10]. MMPs travel along collagen fibrils, degrading these tracks (catalysing the cleavage and subsequent disassembly of collagen) as they travel.  (For more information on collagen fibrils, see the article by Baldwin, Quigley and Kreplak in this issue.) Observations of individual MMP-1 proteins have characterized their motion as biased one-dimensional diffusion, biased by substrate cleavage and hindered by specific interactions with the collagen fibrillar track [11].  Intriguingly, the cleavage sites for MMP are separated by 65 nm along the collagen fibril, a distance approximately an order of magnitude larger than the size of the MMP. There are many open questions regarding how MMPs achieve biased directionality and an extremely high processivity. Recent studies suggest that it is not sufficient to examine simply the MMP, but rather one must consider the collagen fibrillar track and its structure as an integral part of the motility mechanism [12].

An artificial DNA motor that utilizes hydrolysis to induce a burnt-bridges mechanism has been constructed by Bath \textit{et al.} [6]. In their system, the phosphodiester DNA backbone of an anchoring strand of DNA is enzymatically cleaved. The cleavage results in a single strand of DNA preferentially hybridizing the next intact anchor strand, without the option of going back a step.  

Taking a protein-based approach to engineering, the `lawnmower' [13] is a design that exploits a burnt-bridges mechanism, thereby functioning autonomously. Recently constructed, the lawnmower is expected to bias its own motion by enzymatically cleaving peptide substrates along a track. As shown in \textbf{Figure 3}, the lawnmower consists of a quantum dot hub conjugated to multiple protease `blades' via flexible linker molecules. Inspired by the concept of a self-steering lawnmower, this motor was designed to undergo biased motion towards uncleaved substrate.  The proteases serve not only as blades but also as `feet' that first bind, and then cut, their foot-holds (peptides) as they move, thereby effectively biasing motion of the entire complex. The lawnmower is expected to remain on the track as long as the individual foot-binding events occur before complete detachment. The construct is entirely modular. The attached proteases can be switched, the hub size can be increased, and the linker arms connecting the blades and hub can be lengthened or shortened. On a large 2D surface, the motion is expected to resemble a self-avoiding walk, while constraining the track to 1D should result in directed motion. The modular design allows researchers to explore the influence of individual components on motor performance. 

\begin{figure}
	\centering
	\includegraphics[width=0.48\textwidth]{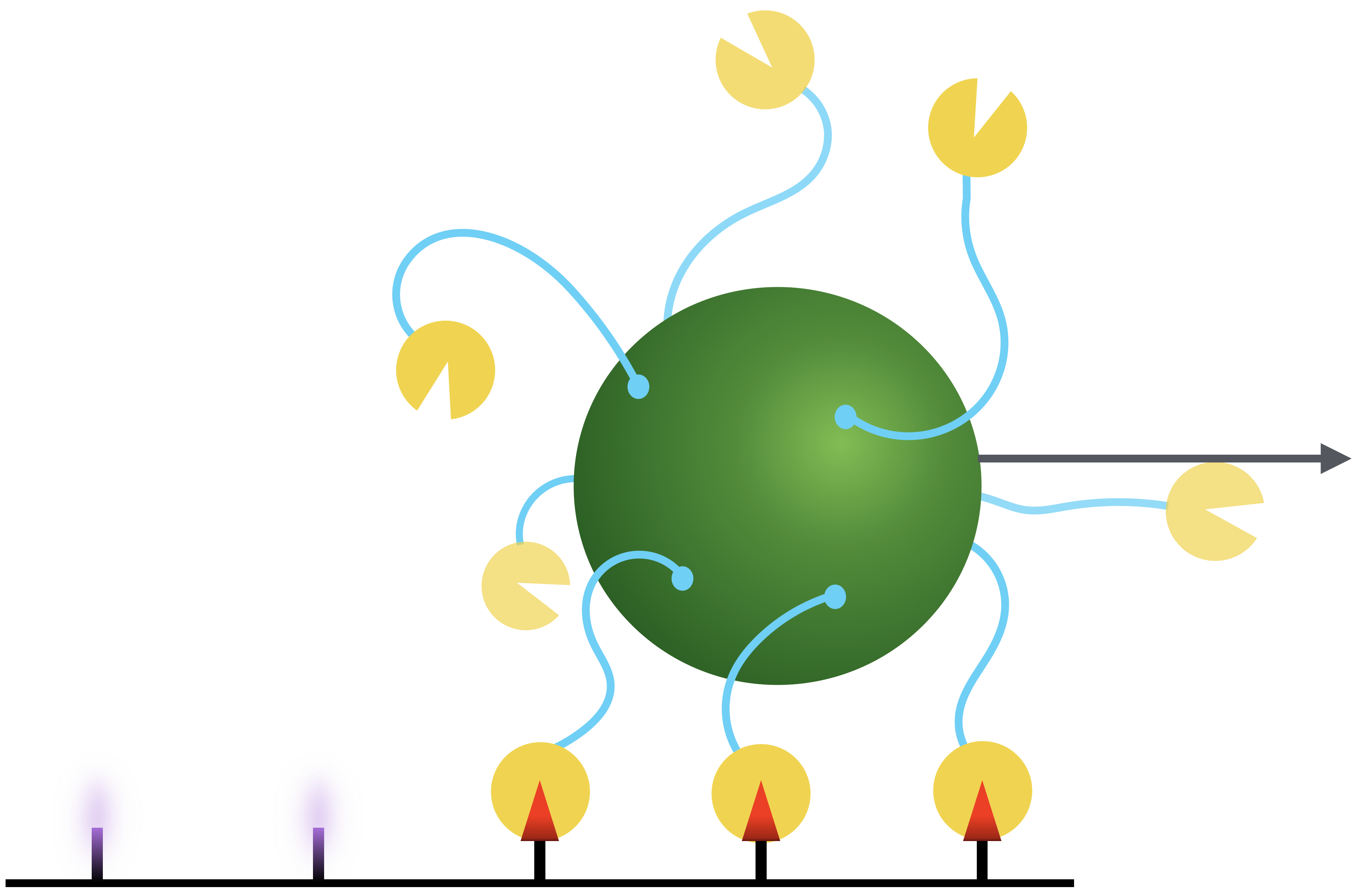}
	\caption{ \textbf{The Lawnmower}. Multiple proteases are coupled to a central quantum dot hub (8 nm in radius), that can be observed via fluorescence. The substrate track is comprised of peptides that contain a recognition sequence for the protease. Following binding, the protease cuts this sequence in two, with one portion diffusing into solution.  Proteases then bind to uncleaved peptides; this biases the motion of the lawnmower in a burnt-bridges fashion. Detection of motion and of track cleavage can be performed using fluorescence: the portion of the peptide released by cleavage contains a quencher, thus resulting in fluorescence emission from the fluorophore remaining on the track-associated peptide stub.  Because of the different spectral properties of the quantum dot and track, the motion of the motor can be tracked and correlated with cleavage of substrate, which provides a read-out of mechanochemical coupling. }
\end{figure}

\section{\label{sec:level1}CONCLUSIONS}

There are many examples of artificial motors being pursued by researchers, each ingeniously designed to walk directionally along a track. In this article we covered examples of biological motors, motile DNA machines, and designs of protein-based synthetic molecular motors. These molecular devices, each in its own way, allow us to learn about the complex biological systems that led to their inspiration. Engineering artificial motors will not only help us understand biological processes, but also holds promise to create new technologies on the nano-scale. For example, DNA devices may have use as sensors for medical applications [14]. Although the field of synthetic nano-scale machinery is still in its infancy, it has great promise for future applications. We are only now realizing the great potential of these devices. There is still a lot to learn.

\section{\label{sec:level1}ACKNOWLEDGEMENTS}

Research on molecular motors in the Forde lab is supported by NSERC. The authors wish to thank David Sivak, Victoria Loosemore and Martin Zuckermann for valuable feedback on this manuscript.

\begin{table}
	\centering
	\caption{Terminology}
	\resizebox{\columnwidth}{!}{%
		\label{my-label}
		\begin{tabular}{l|l}
			\textbf{Ligand}  &          A small molecule that can bind to a protein and alter its function.                                    \\ \hline
			\textbf{ATP}             & Adenosine triphosphate; a source of chemical free energy used to fuel various cellular processes                          \\ \hline
			\textbf{ADP}             & Adenosine diphosphate; a product of ATP hydrolysis.               \\ \hline
			\textbf{Peptide}         & A molecule comprised of covalently linked amino acids.                                                         \\ \hline
			\textbf{Protease}        & A protein that hydrolyzes the covalent bonds that link amino acids together.                                    \\ \hline
			\textbf{Hydrolysis} &   The water-facilitated cleavage of chemical bonds.                                       
		\end{tabular}
	}
\end{table}

\begin{table}
	\centering
	\caption{Terminology}
	\resizebox{\columnwidth}{!}{%
		\label{my-label}
		\begin{tabular}{lp{.5\textwidth}l}{}
			\textbf{Ligand}  &          A small molecule that can bind to a protein and alter its function.                                    \\ \hline
			\textbf{ATP}             & Adenosine triphosphate; a source of chemical free energy used to fuel various cellular processes                          \\ \hline
			\textbf{ADP}             & Adenosine diphosphate; a product of ATP hydrolysis.               \\ \hline
			\textbf{Peptide}         & A molecule comprised of covalently linked amino acids.                                                         \\ \hline
			\textbf{Protease}        & A protein that hydrolyzes the covalent bonds that link amino acids together.                                    \\ \hline
			\textbf{Hydrolysis} &   The water-facilitated cleavage of chemical bonds.                                       
		\end{tabular}
	}
\end{table}

%{lp{.5\textwidth}l}

\section{\label{sec:level1}REFERENCES}

[1] ``Feynman’s office: The last blackboar''. Physics Today, 88 (Feb. 1989).\\

[2] Q. Shao \& Y.Q. Gao. “On the hand-over-hand mechanism of kinesin”. \textit{Proc. Natl. Acad. Sci}., \textbf{103}, 8072-8077 (2006). \\

[3] M. Schliwa \& G. Woehlke. ”Molecular motors”. \textit{Nature}, \textbf{422}, 759-765 (2003). \\

[4] T.L. Blasius, D Cai, GT Jih, CP Toret \& KJ Verhey. “Two binding partners cooperate to activate the molecular motor Kinesin-I”. \textit{J. Cell Biol.}. \textbf{176},11-17 (2007). \\

[5] E. Mayhofer \& J. Howard. “The force generated by a single kinesin molecule against an elastic load”. \textit{Proc. Natl. Acad. Sci.}, \textbf{92}, 574-578 (1995).\\

[6] J. Bath \& A.J. Turberfield. “DNA nanomachines”. \textit{Nat. Nanotechnol.}, \textbf{2}, 275-284 (2007). \\

[7] T. Omabegho, R. Sha \& N.C. Seeman. “A Bipedal DNA Brownian Motor with Coordinated Legs”. \textit{Science}, \textbf{324,} 67-71 (2009). \\

[8] E.H.C. Bromley et al. “The Tumbleweed: Towards a synthetic protein motor”. \textit{HFSP Journal}, \textbf{3}, 204-212  (2009). \\

[9] C.S. Niman et al. “Controlled microfluidic switching in arbitrary time-sequences with low drag”. \textit{Lab on a Chip}, \textbf{13}, 2389-2396 (2013). \\

[10] S. Saffarian, I.E. Collier, B.L. Marmer, E.L. Elson \& G. Goldberg. “Interstitial Collagenase is a Brownian Ratchet Driven by Proteolysis of Collagen”. \textit{Science}, \textbf{306}, 108-111 (2004). \\

[11] S.K. Sarkar, B. Marmer, G. Goldberg \& K.C. Neuman. “Single-Molecule Tracking of Collagenase on Native Type I Collagen Fibrils Reveals Degradation Mechanism”. \textit{Curr. Biol.}, \textbf{22}, 1047-1056 (2012). \\

[12] A. Dittmore, J. Silver, S.K. Sarkar, B. Marmer, G.I. Goldberg \& K.C. Neuman. “Internal strain drives spontaneous periodic buckling in collagen and regulates remodeling”. \textit{Proc. Natl. Acad. Sci.}, \textbf{113}, 8436-8441 (2016). \\

[13] S. Kovacic, L. Samii, P.M.G. Curmi, H. Linke, M.J. Zuckermann \& N.R. Forde. “Design and Construction of the Lawnmower, An Artificial Burnt-Bridges Motor”. \textit{Trans. Nanobiosci.}, \textbf{14}, 305-312 (2015). \\

[14] S. Modi, Swetha M.G., D. Goswami, G.D. Gupta, S. Mayor \& Y. Krishnan. “A DNA nanomachine that maps spatial and temporal pH changes inside living cells”. \textit{Nat. Nanotechnol.}, \textbf{4}, 325-330 (2009). \\

\end{document}